\documentclass[twocolumn, prl]{revtex4}
\usepackage[latin1]{inputenc}
\usepackage{amsmath}
\usepackage{amsfonts}
\usepackage{graphics}
\usepackage{epsfig}
\begin{document}

\title{Quenching of phonon-induced processes in quantum dots due to electron-hole asymmetries}

\author{A. Nysteen, P. Kaer, and J. Mork}
%\email{anys@fotonik.dtu.dk}
\affiliation{DTU Fotonik, Department of Photonics Engineering, Technical University of Denmark, Building 343, 2800 Kgs. Lyngby, Denmark}

\date{\today}

\begin{abstract}
Differences in the confinement of electrons and holes in quantum dots are shown to profoundly impact the magnitude of scattering with acoustic phonons in materials where crystal deformation shifts the conduction and valence band in the same direction. Using an extensive model that includes the non-Markovian nature of the phonon reservoir, we show how the effect may be addressed by photoluminescence excitation spectroscopy of a single quantum dot. We also investigate the implications for cavity QED, i.e. a coupled quantum dot-cavity system, and demonstrate that the phonon scattering may be strongly quenched. The quenching is explained by a balancing between the deformation potential interaction strengths and the carrier confinement and depends on the quantum dot shape. Numerical examples suggest a route towards engineering the phonon scattering. 
\end{abstract}

\maketitle
It is well-known that cavity QED systems undergo dephasing due to interactions with the environment, and that this leads to loss of quantum mechanical coherence between the different states of the system. In particular, solid-state based cavity QED systems such as semiconductor micropillars \cite{Reithmaier2004,Gerard1996,Lermer2012} and photonic crystal cavities \cite{Yoshie2004,Vuckovic2003} are strongly affected by dephasing induced by phonon scattering \cite{Hohenester2010,Ates2009,Hughes2011,Majumdar2011,WilsonRae2002,Muljarov2004,Kaer2010,Gauger2008}. Phonon-induced decoherence thus makes the observation of such effects as vacuum Rabi oscillations \cite{Norris1994} much more difficult than in atomic cavity QED systems and impairs the realization of a scalable solid-state platform for quantum information technology. For instance, the requirement of indistinguishability of subsequent emission events from a single-photon source places strict limitations on the amount of dephasing that can be accepted \cite{KaerLang,Kiraz2004,Bylander2003}. 

In this Letter we investigate the role of carrier confinement on phonon scattering in a semiconductor quantum dot (QD). We employ a comprehensive theoretical model that takes into account the non-Markovian nature of the phonon reservoir \cite{Kaer2010}, thus avoiding the standard approach of describing phonon-induced decoherence by a pure dephasing rate \cite{Krummheuer2002,Auffeves2010,Naesby2008}. Surprisingly, we find that acoustic deformation potential scattering may be completely quenched under certain conditions, depending on the degree of confinement of the involved electronic states. The effect is investigated in detail for two cases.

Firstly, we consider the case of photoluminescence spectroscopy of a single QD and find that the luminescence is suppressed at certain detunings due to a quenching of phonon scattering. For spherical confinement potentials we derive approximate analytical results for the scattering rate. These results show that the effect originates from the difference in the spatial confinement of electrons and holes, usually neglected in theoretical treatments. 

Secondly, we consider a coupled QD-cavity system. We find that the quenching of the phonon scattering strongly affects pure dephasing with consequences e.g. for the indistinguishability of single photon sources. We further expand the description to realistic QD structures, which are analyzed numerically, and the conditions for reducing phonon scattering are established.

Other approaches towards controlling the degree of phonon scattering include the use of phononic bandgap structures for suppressing vibrational modes \cite{Kushwaha1993}. Structures with simultaneous photonic and phononic bandgaps have been discussed theoretically \cite{Maldovan2006} and experimentally realized \cite{Safavi2010} with phononic bandgaps in the GHz-regime. In cavity QED, the most relevant phonons belong to the acoustic branch and have energies in the THz-regime, hence the present phononic band gap structures are primarily of interest for improving the optomechanical coupling \cite{Gavartin2011}. Another approach towards manipulating the electron-phonon interaction has been demonstrated by placing the QD near a surface, thereby changing the phononic dispersion \cite{Krummheuer2005}.

At first we consider a two-level QD with transition energy $\hbar\omega_\textrm{QD}$ illuminated by a CW laser with frequency $\omega_\textrm{L}$ in a standard photoluminescence excitation experiment. We limit ourselves to detunings below $10\,\textrm{meV}$, where the the deformation potential coupling between electrons and longitudinal acoustic phonons constitutes the dominating coupling mechanism \cite{Krummheuer2005_2}. Initially we consider spherical-parabolic confinement potentials, giving ground state wavefunctions
\begin{equation}
\phi_\nu(\mathbf{r})=\pi^{-3/4}w_\nu^{-3/2}\exp\left[-r^2/(2w_\nu^2)\right], \label{eq:wavef}
\end{equation}
with $\nu\in \{e,h\}$ representing the electron and the hole respectively. Numerical results for more realistic QD structures will be presented later. The differences in carrier confinement lead to wavefunctions with different effective widths for electrons ($w_e$) and holes ($w_h$). The equations of motion for the system are derived using a quantum master equation approach \cite{Breuer1999} exploiting recent results \cite{Kaer2010} for the reduced density matrix, $\rho(t)$, where the phononic degrees of freedom have been traced out. The system evolves according to
\begin{eqnarray}
\partial_t\rho(t)=-\textrm{i}\hbar^{-1}[H_\textrm{s},\rho(t)]+S_\textrm{LA}(t)+L(t), \label{eq:reduceddensity}
\end{eqnarray}
where $H_\textrm{s}=\hbar(\omega_\textrm{QD}-\omega_\textrm{L})\sigma_{ee}+\hbar\Omega(\sigma_{eg}+\sigma_{ge})$ is the system Hamiltonian for the QD and the optical excitation \footnote{The system Hamiltonian is similar to the coupled QD-cavity system Hamiltonian in \cite{KaerLang} with $g\rightarrow \Omega$}, corresponding to a Rabi-frequency $\hbar\Omega=10\,\mu\textrm{eV}$. $S_\textrm{LA}(t)$ is a time-local phonon-induced scattering term, and $L(t)$ represents losses included through the Lindblad formalism \cite{Lindblad1976}. The latter accounts, through the rate $\Gamma$, for the decay of the excited QD state due to spontaneous radiation into other modes than the cavity mode as well as non-radiative processes.

\begin{figure}
\includegraphics[width=0.48\textwidth]{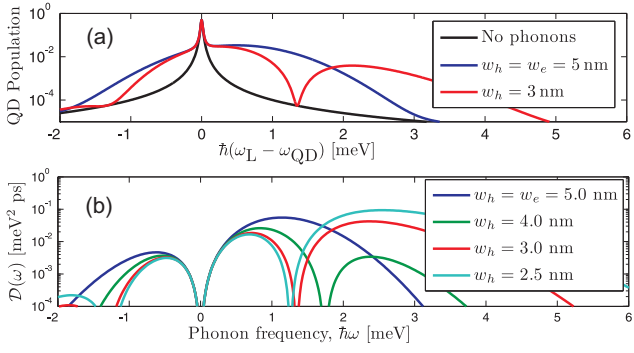}
\caption{\label{PLE}(color online). (\textbf{a}) Stationary population of a bare QD, when excited by a CW field, corresponding to a Rabi-frequency $\hbar\Omega=10\,\mu\textrm{eV}$. The black curve describes the case without phonon-assisted coupling, and the QD decay rate is $\Gamma=1\,\textrm{ns}^{-1}$. (\textbf{b}) The effective phonon spectrum for spherical wavefunctions for varying $w_h$ and $w_e$. For both plots, $w_e^3+w_h^3$ is kept constant, and $T=4\,\textrm{K}$.}
\end{figure}

Solving Eq. (\ref{eq:reduceddensity}) numerically \footnote{We consider parameters for GaAs, with deformation potentials $D_e=-14.6\,\textrm{eV}$, $D_h=-4.8\,\textrm{eV}$ \cite{Hohenester2010,Krummheuer2002}, $c_l=5110\,\textrm{m/s}$, and $d=5370\,\textrm{kg/m}^3$.}, we calculate the stationary population of an InAs QD embedded in GaAs in depending on the detuning between the laser and the QD ground state resonance, see Fig. \ref{PLE}a. Comparing the black and blue curve in Fig. \ref{PLE}a shows that the QD indeed, as is well-known \cite{Hughes2011_PrX,HughesNye}, may be populated by a phonon-assisted process, even when the laser and the QD are off-resonant. However, when the widths of the electron and hole wavefunction differ, as for the red curve, the QD population and thus the photoluminescence intensity, is suppressed at a specific detuning. As we shall show, this implies that the phonon scattering is quenched, which introduces a new and interesting approach for manipulating the electron-phonon interaction. 

The strength of the carrier-phonon scattering is quantified by the effective phonon spectrum, $\mathcal{D}(\omega)$, describing how the phonon modes interact with the QD at a given temperature. Positive phonon frequencies refer to a phonon of energy $\hbar\omega$ being emitted into the surroundings, whereas negative phonon frequencies indicates the absorption of a phonon of energy $\hbar|\omega|$ from the surrounding environment. $\mathcal{D}(\omega)$ is calculated as the real part of the Fourier transform of the phonon reservoir correlation function $\mathcal{D}(t)$ \cite{Kaer2010}. We have
\begin{equation}
\mathcal{D}(\omega)=\pi\sum_\mathbf{k}|M^\mathbf{k}|^2\bigg[n_\mathbf{k}\delta(\omega+\omega_\mathbf{k})+(n_\mathbf{k}+1)\delta(\omega-\omega_\mathbf{k})\bigg],
\end{equation}
where $\mathbf{k}$ is the phonon wavevector, $n_\mathbf{k}=1/[\exp(\hbar\omega_\mathbf{k}/k_BT)-1]$ is the average thermal occupation number of the phonon mode $\mathbf{k}$ at temperature $T$. $M^\mathbf{k}=M^\mathbf{k}_{ee}-M^\mathbf{k}_{hh}$ is the effective electron/hole-phonon coupling matrix element with
\begin{equation}
M^\mathbf{k}_{\nu\nu}=D_\nu\sqrt{\frac{\hbar k}{2dc_lV}}\mathcal{F}_\nu(\mathbf{k}). \label{eq:intmat}
\end{equation}
The electronic form factor, 
\begin{equation}
\mathcal{F}_\nu(\mathbf{k})=\int\,\textrm{d}\mathbf{r}|\phi_\nu(\mathbf{r})|^2 \textrm{e}^{\textrm{i}\mathbf{k}\cdot\mathbf{r}}, \label{eq:formfac}
\end{equation}
describes how a phonon with wavevector $\mathbf{k}$ interacts with the confined carriers, governed by the overlap between the carrier and phonon wavefunctions. We assume bulk phonons with a linear dispersion relation, $\omega_\mathbf{k}=c_l|\mathbf{k}|$. $D_e$ ($D_h$) is the deformation potential constant of a conduction band electron (valence band hole), $c_l$ is the velocity of longitudinal acoustic waves, $d$ is the bulk density, and $V$ is the phonon quantization volume.

For spherical wavefunctions, cf. Eq. (\ref{eq:wavef}), the effective phonon spectrum reduces to
\begin{eqnarray}
\mathcal{D}(\omega)=\frac{\hbar}{4\pi dc_l^5}\frac{\omega^3}{1-\text{e}^{-\beta \hbar\omega}} \;\;\;\;\;\;\;\;\;\;\;\;\;\;\;\;\;\;\;\;\;\;\;\;\;\;\;\;\; \nonumber\\ \;\;\;\;\;\;\;\;\;\;\;\;\;\;\; \times \left[D_e\text{e}^{-\omega^2w_e^2/(4c_l^2)}-D_h\text{e}^{-\omega^2w_h^2/(4c_l^2)}\right]^2, \label{eq:dspher}
\end{eqnarray}
and is shown in Fig. \ref{PLE}b. The magnitude of $\mathcal{D}(\omega)$ is generally smaller for $\omega<0$ compared to $\omega>0$ due to the low probability of thermally excited phonons. From Eq. (\ref{eq:formfac}) we note that a spatially narrow wavefuntion is wide in $\mathbf{k}$-space and thus interacts with many phonon modes, explaining why $\mathcal{D}(\omega)$ broadens and increases in magnitude as $w_h$ decreases. For $T>0$, $\mathcal{D}(\omega)$ has three zeros at
\begin{equation}
\omega=0, \;\;\;\omega^2=\frac{4c_l^2}{w_e^2-w_h^2}\ln\left(\frac{D_e}{D_h}\right). \label{eq:dspectr}
\end{equation}
In materials where $D_e/D_h>0$, like GaAs \cite{Hohenester2010,Krummheuer2002}, dips thus appear in the effective phonon spectrum for non-zero phonon energies for $w_e\neq w_h$. The zeros in the effective phonon spectrum appear exactly when $|M^\mathbf{k}|\propto |D_e\mathcal{F}_e(k=\omega/c_l)-D_h\mathcal{F}_h(k=\omega/c_l)|=0$ $\Rightarrow D_e\mathcal{F}_e(k=\omega/c_l)=D_h\mathcal{F}_h(k=\omega/c_l)$. 

Based on this we can explain the quenching of phonon-induced processes. The deformation potential interaction occurs in general due to the different values of the deformation potential constant in the conduction and valence band. However, for a specific detuning, this can be compensated by a difference in the confinement of the electron and the hole through the form factor in Eq. (\ref{eq:formfac}). In the case of identical electron and hole wavefunctions, $w_e=w_h$, the zeros appear at infinity, and the effect is not apparent. 

\begin{figure}
\includegraphics[width=0.45\textwidth]{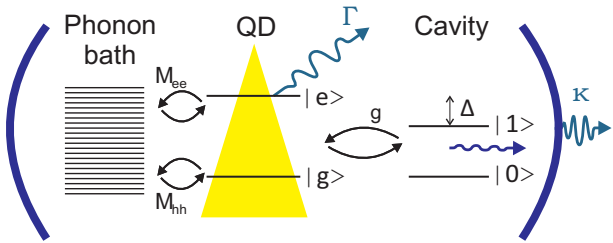}
\caption{\label{figSketch}(color online). Coupled cavity--QD system interacting with a phonon reservoir. The optical coupling strength is $g$, the QD--cavity detuning is $\Delta$ (positive for the case shown), and $M_{ee/hh}$ are electron/hole--phonon coupling matrix elements. $\Gamma$ is the QD population decay rate, and $\kappa$ is the leakage rate from the optical cavity.}
\end{figure}

The quenching of phonon processes has interesting consequences for cavity QED. Instead of illuminating the QD with a laser, the QD is now placed inside a single-mode optical cavity, detuned by $\Delta=\omega_\textrm{QD}-\omega_\textrm{cav}$ from the cavity resonance, see Fig. \ref{figSketch}. In this case, $H_\textrm{s}$ in Eq. (\ref{eq:reduceddensity}) is replaced by the combined QD-cavity system Hamiltonian (corresponding to replacing $\Omega$ with the optical interaction strength $g$ with the cavity mode), see e.g. \cite{KaerLang}. Furthermore, the escape of photons through the cavity is included via a rate $\kappa$ in the Lindblad term $L(t)$ in Eq. (\ref{eq:reduceddensity}).

\begin{figure}
\includegraphics[width=0.48\textwidth]{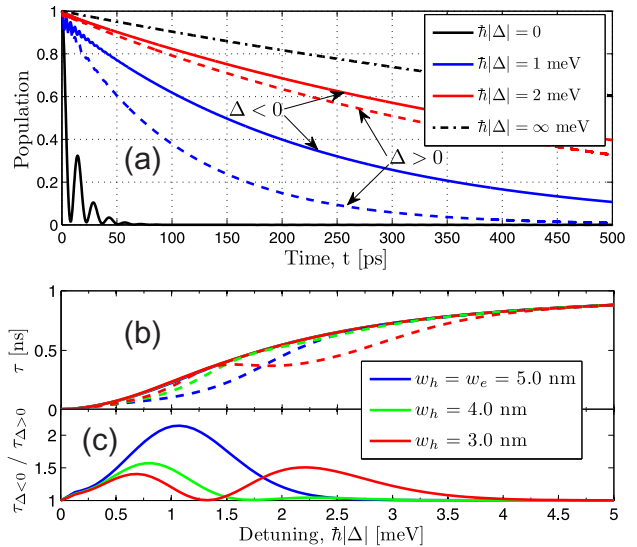}
\caption{\label{fig1}(color online). (\textbf{a}) Time evolution of the population of an initially excited QD for different detunings. The solid (dashed) curve is for negative (positive) detuning, and the electron and the hole are equally confined, $w_e=w_h=5\,\textrm{nm}$. The temperature is $T = 4 \,\textrm{K}$, and we use $\hbar g=150\,\mu\textrm{eV}$, $\Gamma=1\,\textrm{ns}^{-1}$, and $\hbar\kappa=100\,\mu\textrm{eV}$. (\textbf{b}) Lifetimes, $\tau$, of the excited QD state plotted versus detuning for different values of $w_e$ and $w_h$, keeping the volume parameter $w_e^3+w_h^3$ constant. The solid (dashed) curves indicate negative (positive) detunings. (\textbf{c}) The degree of asymmetry versus detuning.} 
\end{figure}

By solving Eq. (\ref{eq:reduceddensity}), we obtain the QD decay curves shown in Fig. \ref{fig1}a. The structure of the phonon bath results in an asymmetry, expressed in the possibility of the QD to couple to a red-tuned (lower energy, $\Delta>0$) cavity by the emission of an acoustic phonon, but a lack of coupling to a blue-tuned (higher energy, $\Delta<0$) cavity by phonon absorption at low temperatures, where the population of thermally excited phonons is low \cite{Hohenester2010,Kaer2010}. This is in accordance with $\mathcal{D}(\omega)$ in Fig. \ref{PLE}b. In the limit of $|\Delta|\rightarrow\infty$, the QD and the cavity decouple, and the QD decays with the rate $\Gamma$. By single-exponential fits to the QD decay curves, the lifetime of the excited QD state is calculated, see Fig. \ref{fig1}b. From this we determine the lifetime ratio $\tau_{\Delta<0}/\tau_{\Delta>0}$, see Fig. \ref{fig1}c. Surprisingly, we observe that when $w_e\neq w_h$, a non-zero QD--cavity detuning exists, where no lifetime asymmetry is present. 

To examine the physical origin of the lifetime asymmetry, we consider the total decay rate of the excited QD state in the large detuning limit $\Delta\gg g$ \cite{KaerLang},
\begin{equation}
\Gamma_\text{tot}\approx\Gamma+2g^2\frac{\gamma_\text{tot}}{\gamma_\text{tot}^2+\Delta^2}\left[1+\frac{1}{\hbar^2\gamma_\text{tot}}\mathcal{D}(\omega=\Delta)\right], \label{eq:PersExp}
\end{equation}
with $\gamma_\text{tot}=(\Gamma+\kappa)/2$. The total QD decay rate, $\Gamma_\textrm{tot}$, has two contributions besides the decay rate $\Gamma$. The first term in the square brackets represents the usual Purcell enhancement rate, while the second term represents the decay of the QD through the cavity by the simultaneous emission/absorption of a phonon. Thus the behaviour of $\mathcal{D}(\omega)$ translates directly into the behaviour of the lifetime ratio plot in Fig. \ref{fig1}c, such that non-zero QD-cavity detunings exist, where phonons do not affect the lifetime of the excited QD state.

\begin{figure}
\includegraphics[width=0.49\textwidth]{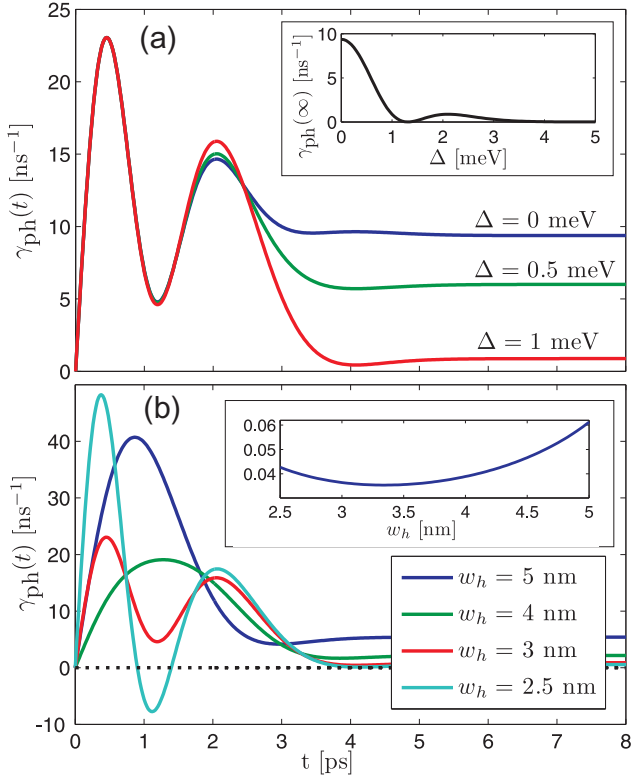}
\caption{\label{fig444_ny}(color online). (\textbf{a}) Time evolution of phonon-induced dephasing rate for different QD-cavity detunings. Inset show the long-time value of the dephasing rate versus detunings, $w_h=3\,\textrm{nm}$, $w_e=6\,\textrm{nm}$, and $T=4\,\textrm{K}$. (\textbf{b}) Time evolution of phonon-induced dephasing rate for different wavefunction widths, keeping $w_e^3+w_h^3$ constant, and $\Delta=1\,\textrm{meV}$. The inset quantifies the short-time scattering by showing the integrated value of $\gamma_\textrm{ph}(t)$ from $t=0$ to 3 ps for different wavefunction widths.}
\end{figure}

The suppression of the effective phonon density is expected to affect not only the QD lifetime, but all phonon-induced effects such as pure dephasing. The degree of single-photon indistinguishability \cite{Kiraz2004} is typically quantified as the ratio between the coherence time and lifetime of the emitter \cite{Bylander2003}. For our system, this ratio is $\Gamma_\textrm{tot}/(\Gamma_\textrm{tot}+2\gamma_\textrm{ph})$, with $\gamma_\textrm{ph}$ being the pure dephasing contribution from the phonons. This hints at the importance of having a high decay rate compared to the dephasing rate.

The phonon-induced pure dephasing rate \footnote{$\gamma_\textrm{ph}$ corresponds to $\textrm{Re}\{\gamma_{12}\}$ in \cite{Kaer2010}.} is calculated from the phonon bath correlation function,
\begin{eqnarray}
\gamma_\textrm{ph}(t)=\textrm{Re}\bigg\{\hbar^{-2}(1-K)\int_0^t\mathrm{d}t'\,\mathcal{D}(t')\;\;\;\;\;\;\;\;\;\;\;\;\;\;\;\;\;\;\;\;\;\nonumber\\
\;\;\;\;\;\;\;\;\;\;\;\;\;\;\;+\hbar^{-2}K\int_0^t\mathrm{d}t'\,\cos\left(t'\sqrt{\Delta^2+4g^2}\right)\mathcal{D}(t')\bigg\}, \label{eq:inds}
\end{eqnarray}
where $K$ is a parameter depending solely on the ratio $\Delta/g$. For bulk phonons, $\mathcal{D}(t)$ decays within 5 ps \cite{KaerLang}. As seen from Fig. \ref{fig444_ny}a, the initial behaviour of $\gamma_\textrm{ph}(t)$ is governed by the bare QD-phonon coupling, whereas the long-time value depends on the cavity detuning. On the short time scale, the phonons participate in non-energy conserving virtual processes. Thus the electrons interact with the full phonon spectrum $\mathcal{D}(\omega)$. The long-time ($t\rightarrow\infty$) value of $\gamma_\textrm{ph}(t)$, on the other hand, implies a Fourier transform of $\mathcal{D}(t)$ and depends only on $\mathcal{D}(\omega=0)$ and $\mathcal{D}(\omega=\pm\sqrt{\Delta^2+4g^2})$, where the former is zero. % J: Corresponding to the polariton branches

The phonon-induced dephasing of the QD--cavity system may be reduced by minimizing both the short-time and long-time scattering. The former may be achieved by reducing the overall amplitude of $\mathcal{D}(\omega)$, as illustrated in Fig. \ref{fig444_ny}b for $w_h\sim 3.4\,\textrm{nm}$, compare with Fig. \ref{PLE}b. The reduction of long-time scattering is obtained by using a QD--cavity detuning such that $\sqrt{\Delta^2+4g^2}$ coincides with a dip in $\mathcal{D}(\omega)$, as seen from the inset in Fig. \ref{fig444_ny}a. It turns out that for the indistinguishability of single photons, the reduction of the overall amplitude is the most important \cite{KaerIndist}.

QD structures realized by epitaxial growth are not spherically symmetric \cite{Stobbe2009}, and to investigate the quenching for more realistic structures, we model truncated conical QD structures by solving the one-band effective mass Schrödinger equation using the Finite Element Method \cite{Melnik2004}.  

\begin{figure}
\includegraphics[width=0.485\textwidth]{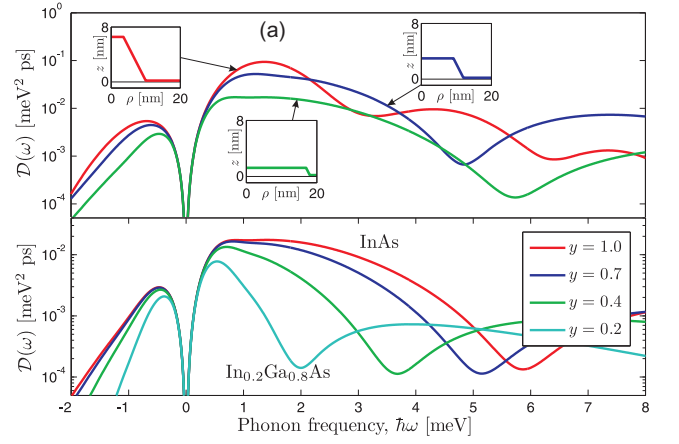}
\caption{\label{fig6}(color online). (\textbf{a}) The effective phonon spectrum calculated from the FEM wavefunctions of the three QDs shown in the insets with matching colors, $T=4\,\textrm{K}$. All QDs have constant volume and side wall slope. (\textbf{b}) The effective phonon spectrum for the shallow dot in (a) for different material compositions, In$_y$Ga$_{1-y}$As, of the QD.} 
\end{figure}

The effective phonon spectrum for the FEM wavefunctions is shown in Fig. \ref{fig6}a for an InAs QD and wetting layer embedded in a GaAs barrier \footnote{We use $E_{g,\text{GaAs}}=1.424\,\text{eV}$, $E_{g,\text{InAs}}=0.359\,\text{eV}$, $m^*_{c,\text{GaAs}}=0.0665m_0$, $m^*_{c,\text{InAs}}=0.027m_0$, $m^*_{v,\text{GaAs}}=0.38m_0$, and $m^*_{v,\text{InAs}}=0.34m_0$.}. The overall amplitude of the spectrum is the largest, when the carriers are strongly confined in two directions, i.e. for the tallest dot, compared to the shallow dot where the carriers are only strongly confined in one direction. These results are qualitatively explained by the form factor in Eq. (\ref{eq:formfac}).  %J: KOMMENTER PÅ MULIGHEDEN FOR AT HAVE FLERE DIPS FOR $\omega>0$

%Using Eq. (\ref{eq:dspectr}) the position of the dip in $\mathcal{D}(\omega)$ may be estimated, assuming that either the $z$- or the $\rho$-direction is dominating, i.e. $w_e=l_{e,\rho}$ and $w_h=l_{g,\rho}$ , or $w_e=l_{e,z}$ and $w_h=l_{g,z}$. The estimated zeros are shown in Fig. \ref{fig6}, demonstrating clearly that the dominating direction is the one, where the carriers are confined the most, when evaluating the dip position from the spherical approximation.

To examine the role of material composition, we consider an In$_y$Ga$_{1-y}$As QD, where the amount of gallium in the QD and wetting layer is varied. By increasing the gallium concentration the band offsets shrink \cite{Vurgaftman2001} and the effective mass changes, %\footnote{LINEAR INTERPOLATION ODER WAS?} 
which results in the phonon spectrum in Fig. \ref{fig6}b. Thus, $w_e$ and $w_h$ both increase, resulting in a lower overall amplitude of the effective phonon spectrum. Due to the heavier hole mass, the asymmetry between the electron and hole wavefunctions also increases, moving the dip in the spectrum towards lower frequencies, as predicted by the spherical model in Eq. (\ref{eq:dspectr}) when $w_e^2-w_h^2$ increases. The position of the dips may in this way be changed depending on the growth parameters of the QD. For small gallium concentrations, the confinement energies may become comparable to the Coulomb energy, in which case the exitonic nature of the electron--hole pair will have to be taken into account.

In conclusion, through the analysis of a comprehensive model, we have learned that electron-phonon scattering may be suppressed due to differences in the spatial confinement of electrons and holes. We suggest that the effect may be measured by photoluminescence excitation spectroscopy, but also that it should strongly affect the decoherence properties of cavity QED systems. We expect this phonon-quenching to be important for the engineering of the phononic properties of devices for quantum information technology, like single-photon sources and switches.

The authors acknowledge helpful discussions with Jørn M. Hvam and financial support from the Villum foundation via the NATEC Centre of Excellence.

\end{document}